%% file: eprint_dpf2015.tex
\documentclass[12pt]{article}
\usepackage{graphicx}

\newcommand\pubnumber{DPF2015-393}
\newcommand\pubdate{\today}

\def\fermilab{Fermi National Accelerator Laboratory\\
Batavia, IL 60510, USA}

\def\Title#1{\begin{center} {\Large #1 } \end{center}}
\def\Author#1{\begin{center}{ \sc #1} \end{center}}
\def\Address#1{\begin{center}{ \it #1} \end{center}}

\newcommand\pubblock{\rightline{\begin{tabular}{l} \pubnumber\\
         \pubdate  \end{tabular}}}
\newenvironment{Abstract}{\begin{quotation}  }{\end{quotation}}
\newenvironment{Presented}{\begin{quotation} \begin{center} 
             PRESENTED AT\end{center}\bigskip 
      \begin{center}\begin{large}}{\end{large}\end{center} \end{quotation}}

\input econfmacros.tex

\begin{document}
\begin{titlepage}
\pubblock

\vfill
\Title{Recent QCD results from the Tevatron}
\vfill
\Author{ Costas Vellidis}
\Address{\fermilab}
\vfill
\begin{Abstract}
Four years after the shutdown of the Tevatron proton-antiproton collider,
the two Tevatron experiments, CDF and DZero, continue producing important
results that test the theory of the strong interaction, Quantum Chromodynamics
(QCD). The experiments exploit the advantages of the data sample acquired
during the Tevatron Run II, stemming from the unique $p\overline{p}$ initial
state, the clean environment at the relatively low Tevatron instantaneous
luminosities, and the good understanding of the data sample after many years
of calibrations and optimizations. A summary of results using the full
integrated luminosity is presented, focusing on measurements of prompt photon
production, weak boson production associated with jets, and non-perturbative
QCD processes.
\end{Abstract}
\vfill
\begin{Presented}
DPF 2015\\
The Meeting of the American Physical Society\\
Division of Particles and Fields\\
Ann Arbor, Michigan, August 4--8, 2015\\
\end{Presented}
\vfill
\end{titlepage}
\def\thefootnote{\fnsymbol{footnote}}
\setcounter{footnote}{0}

\section{Introduction}

The Tevatron proton-antiproton collider at the Fermi National Accelerator
Laboratory (Fermilab) operated at a collision energy of 1.96 TeV (980 GeV
per beam) from 2002 through 2011 (Run II) and was the highest-energy machine
in the World until 2010, when the LHC proton-proton collider at CERN started
operations. At the end of Run II, two short runs were performed at collision
energies of 300 and 900 GeV. During Run II, the collider delivered an
integrated luminosity of approximately 12 fm$^{-1}$. The two experiments,
CDF and DZero, acquired data corresponding to an integrated luminosity of
approximately 10 fm$^{-1}$ each. The two multi-purpose detectors consisted
of high-resolution tracker, calorimeter, and muon detection compartments.
Both were equipped with silicon detectors allowing for precise vertex
reconstruction and thus for heavy-flavor jet tagging (``b-tagging'') with
secondary vertex algorithms. They covered pseudorapidities of up to
$\vert\eta\vert\sim 2$ for leptons, $\vert\eta\vert\sim 4$ for light-flavor
jets, and $\vert\eta\vert\sim 2$ for heavy-flavor jets. The two Collaborations
have been continuously analyzing the acquired data for measurements in all
areas of particle physics, including Standard Model (SM) tests, precision
measurements of SM parameters, and searches for new physics
\cite{publicresults}.

\begin{figure}[htb]
\centering
\includegraphics[width=6in]{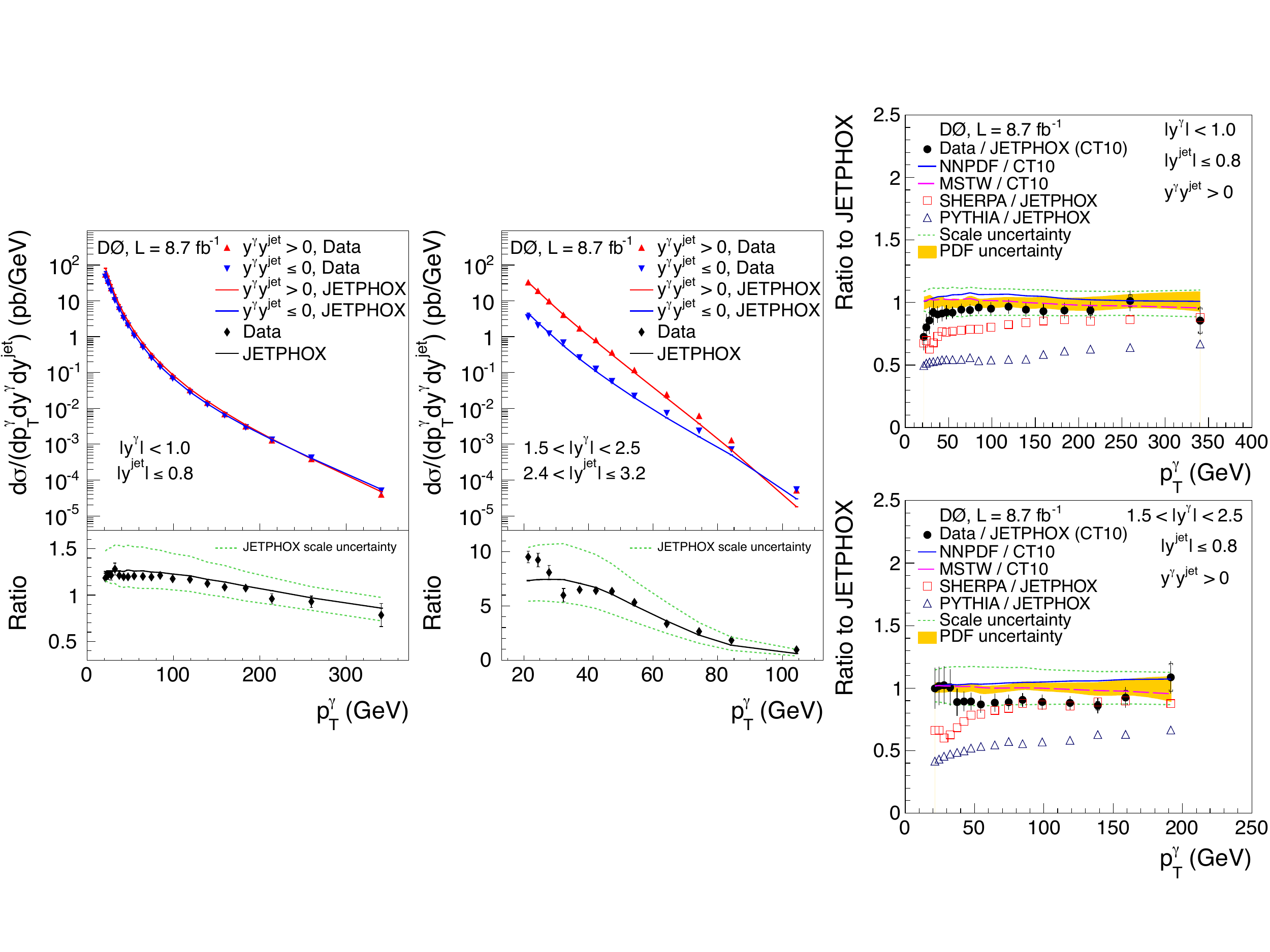}
\vspace{-0.5in}
\caption{Measured and predicted photon$+$jet production cross section
as a function of the photon transverse momentum for central (left) and
forward (middle) photons. The bottom panels of the left and middle plots
show the ratios of the measured cross section to the {\sc jetphox} prediction.
The plots on the right show the ratios of the measured cross section and
various other predictions to the {\sc jetphox} prediction.}
\label{fig:d0phojet}
\end{figure}

Studies of the strong interaction using the Tevatron data are complementary
to those using LHC data in the same kinematic regions. The main reason is
the different initial state. At the Tevatron, the $p\overline{p}$ initial
state favors contributions from processes initiated by valence quarks. For
example, in leading order in the strong coupling, vector bosons are mostly
produced via $q\overline{q}$ annihilation of valence quarks. This is an
advantage, since the parton density functions (PDF) for valence quarks have
smaller uncertainties than those for gluons and sea quarks. At the LHC, on
the other hand, the $pp$ initial state and the higher collision energy favor
contributions from processes initiated by gluons and sea quarks. For example,
in leading order, vector bosons are produced either via $q\overline{q}$
annihilation where the antiquark comes from the sea or via Compton-like
quark-gluon scattering, while gluon fusion, although of higher order, is also
important due to the high gluon luminosity at the LHC energies.

\begin{figure}[htb]
\centering
\includegraphics[width=5in]{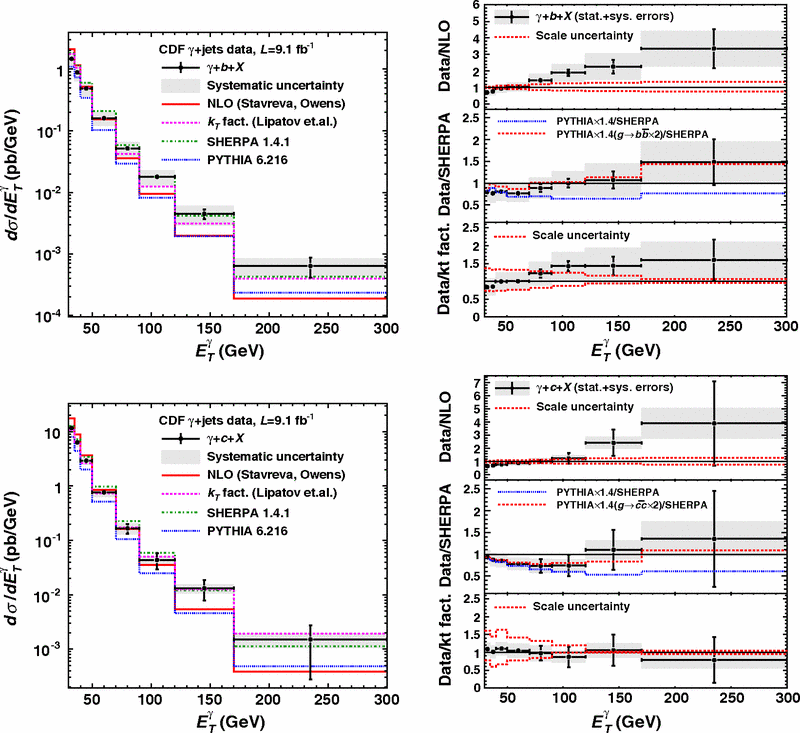}
\caption{Measured and predicted cross section as a function of the photon
transverse energy for photon$+$b quark (top) and photon$+$c quark production
(bottom). The right plots show the ratios of the measured cross section to
various predictions.}
\label{fig:cdfphohf}
\end{figure}

Calculations based on QCD theory can be generally grouped in three types:
perturbative calculations of parton-level processes at a fixed order in
the strong coupling; parton-shower calculations including hadronization
models of the partons; and calculations of parton-level processes resummed
to all orders at some logarithmic accuracy over soft gluon emission, and then
matched to a fixed-order calculation of the hard scattering process.
Fixed-order calculations at next-to-leading order (NLO) are available for
all V$+$jets production process, where V stands for a vector boson (photon,
Z or W). An undergoing effort for full next-to-next-to-leading order (NNLO)
calculations started with the diphoton production. Parton-shower calculations,
effectively performing resummation at leading logarithmic accuracy, started
long ago with leading-order (LO) matrix elements. They now have advanced to
include higher-order matrix elements at the ``tree'' level (represented by
diagrams without loops) and full NLO matrix elements matched to showering
and hadronization models. Calculations of this type provide realistic event
representations at the particle (hadron) level, typically including an
underlying event model for additional parton activity in a hadonic collision,
besides the hard parton scattering, and are thus suitable for simulations
with detailed detector models. Finally, there exists an extensive phenomenology
for non-perturbative QCD, including long-range matrix elements, diffractive
models, and underlying event models.

\begin{figure}[htb]
\centering
\includegraphics[width=6in]{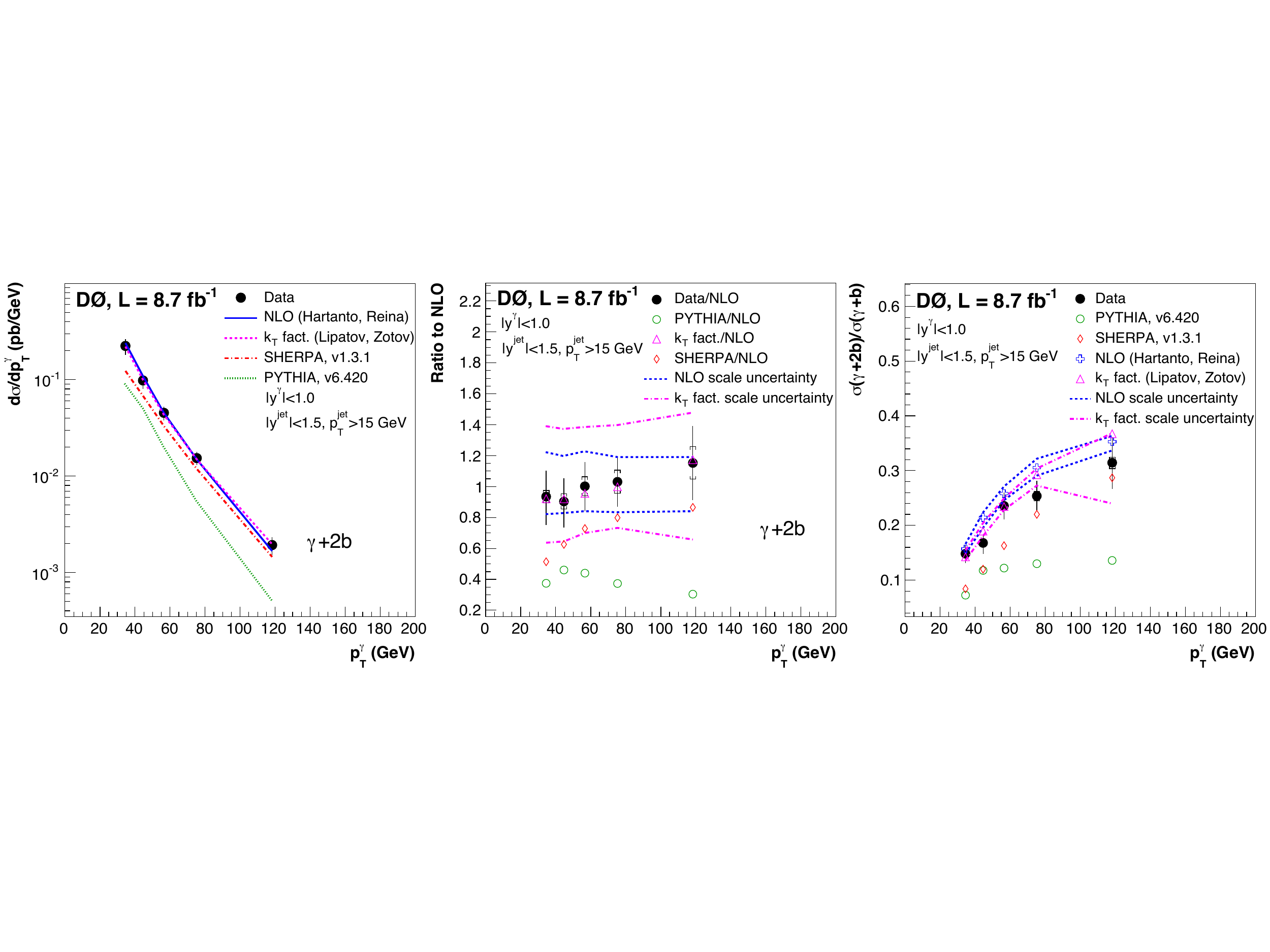}
\vspace{-1.5in}
\caption{Measured and predicted photon$+$2b quark production cross section
(left), and ratios of the measured cross section to the NLO prediction (middle)
and to the measured photon$+$b quark production cross section (right) as a
function of the photon transverse momentum.}
\label{fig:d0pho2b}
\end{figure}

\section{Prompt photon production}

The DZero Collaboration measured the cross section for prompt isolated photon
production in association with light-flavor jets \cite{d0phojet}. The cross
section for photon plus one jet production, differential in the photon
transverse momentum $p_T^\gamma$, is shown in Figure~\ref{fig:d0phojet}. It
is presented for central photons and jets (with absolute rapidities
$\vert y^\gamma\vert<1.0$ and $\vert y^{\rm jet}\vert\le 0.8$) in the left
window and for forward photons and jets (with $1.5<\vert y^\gamma\vert<2.5$
and $2.4<\vert y^{\rm jet}\vert\le 3.2$) in the middle window. In both regions,
the cases of the photon and jet emitted close to each other
($y^\gamma y^{\rm jet}>0$) and away from each other ($y^\gamma y^{\rm jet}\le 0$)
are considered separately, as they probe different parton $x$ ranges. The
difference is more important in the forward region. The measurements are
compared with predictions of the fixed-order NLO program {\sc jetphox}
\cite{jetphox}. This program also implements a phenomenological fragmentation
model handling the collinear photon emission from final state partons, which
causes a singularity in the fixed-order cross section calculation. The
description of the data by {\sc jetphox} is generally successful within
uncertainties. In the right window, the ratio of the data to the {\sc jetphox}
prediction for central jets and central (top) or forward (bottom) photons is
compared with the ratios of tree-level parton-shower calculations to
{\sc jetphox}. The sensitivity to the PDF choice is also examined. While the
{\sc jetphox} prediction agrees with the data within uncertainties, except
in the very low $p_T^\gamma$ region, the parton-shower calculations generally
underestimate the data.

\begin{figure}[htb]
\centering
\includegraphics[width=2.5in]{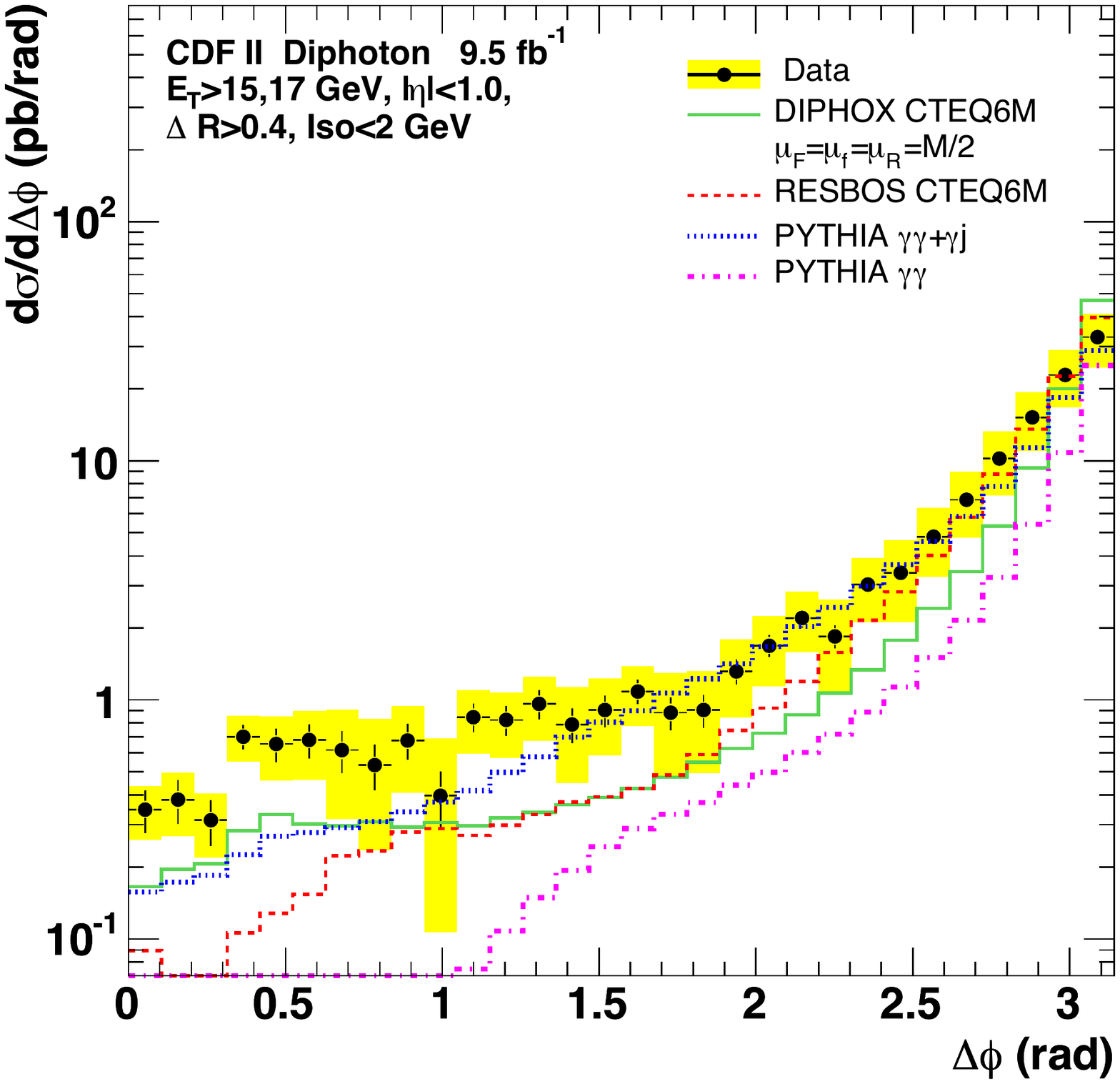}
\includegraphics[width=2.5in]{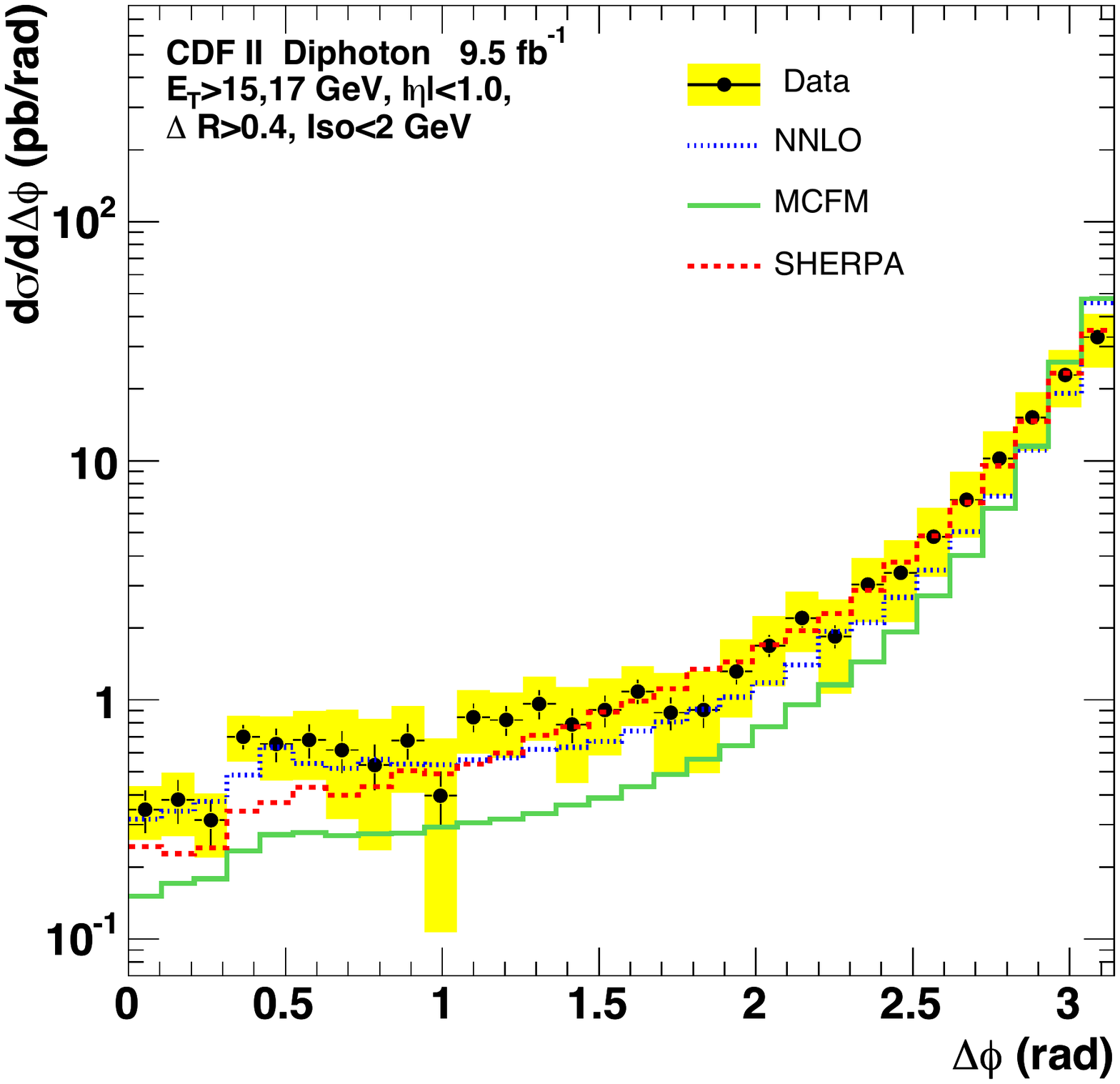}
\caption{Measured and predicted cross section for photon pair production
as a function of the azimuthal angle between the two photons in the event.}
\label{fig:cdfdipho}
\end{figure}

The CDF Collaboration measured the cross section for prompt isolated central
photon production in association with a heavy-flavor quark \cite{cdfphohf},
charm or bottom. The cross section, differential in the photon transverse
energy $E_T^\gamma$, is shown in Figure~\ref{fig:cdfphohf}. In the left window,
the absolute cross section for photon$+$b quark (top) and photon$+$c quark
production (bottom) is displayed. In the right window, the corresponding
ratios of the measured cross section to various theoretical predictions are
displayed, for better visualization of the comparisons between data and theory.
The analytically resummed ``k$_{\rm T}$-factorization'' calculation \cite{ktfac}
and the parton-shower {\sc sherpa} calculation \cite{sherpa} agree with the
data within uncertainties, whereas the fixed-order NLO calculation fails to
describe the data.

\begin{figure}[htb]
\centering
\includegraphics[width=2.5in]{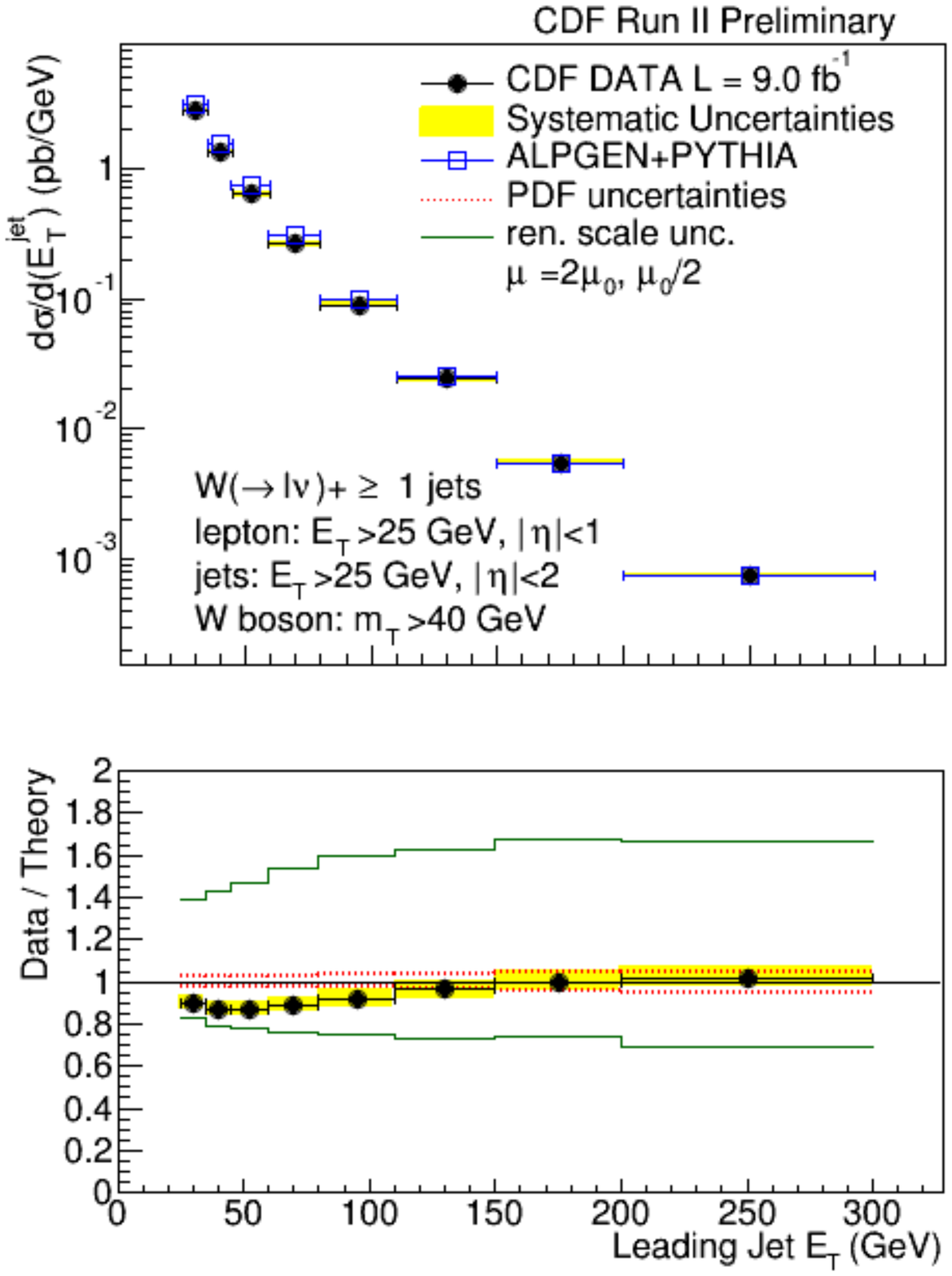}
\includegraphics[width=2.5in]{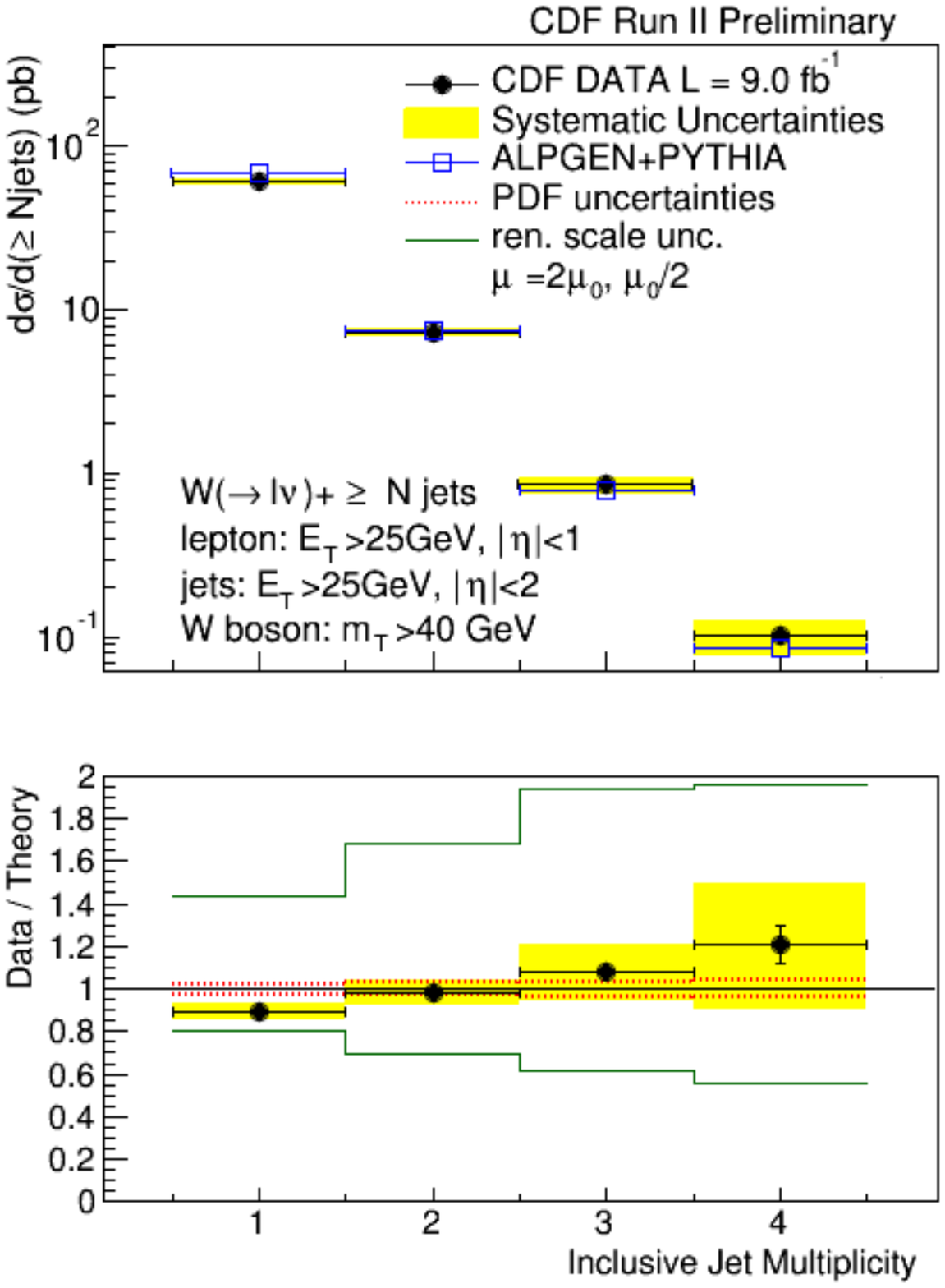}
\caption{Measured and predicted cross section for W$+$jets production as a
function of the leading jet transverse energy (left) and of the number of
jets in the event (right). The bottom panels show the corresponding ratios
of the measurement to the prediction.}
\label{fig:cdfwjet}
\end{figure}

An extension of the previous measurement comes from DZero, who measured
the cross section for prompt isolated central photon production in association
with two b quarks \cite{d0pho2b}. Figure~\ref{fig:d0pho2b} shows the cross
section, differential in $p_T^\gamma$, measured and predicted by various
calculations. The cross section is displayed in the left window. The ratios
of measured and predicted cross sections to the fixed-order NLO prediction
are displayed in the middle window. The ratios of the photon$+$2b cross section
to the photon$+$b cross section \cite{d0phohf} for the data and the various
predictions are displayed in the right window. In this case, parton-shower
calculations fail to describe the data, whereas the fixed-order NLO and
k$_{\rm T}$-factorization calculations agree with the data within uncertainties.

\begin{figure}[htb]
\centering
\includegraphics[width=2.5in]{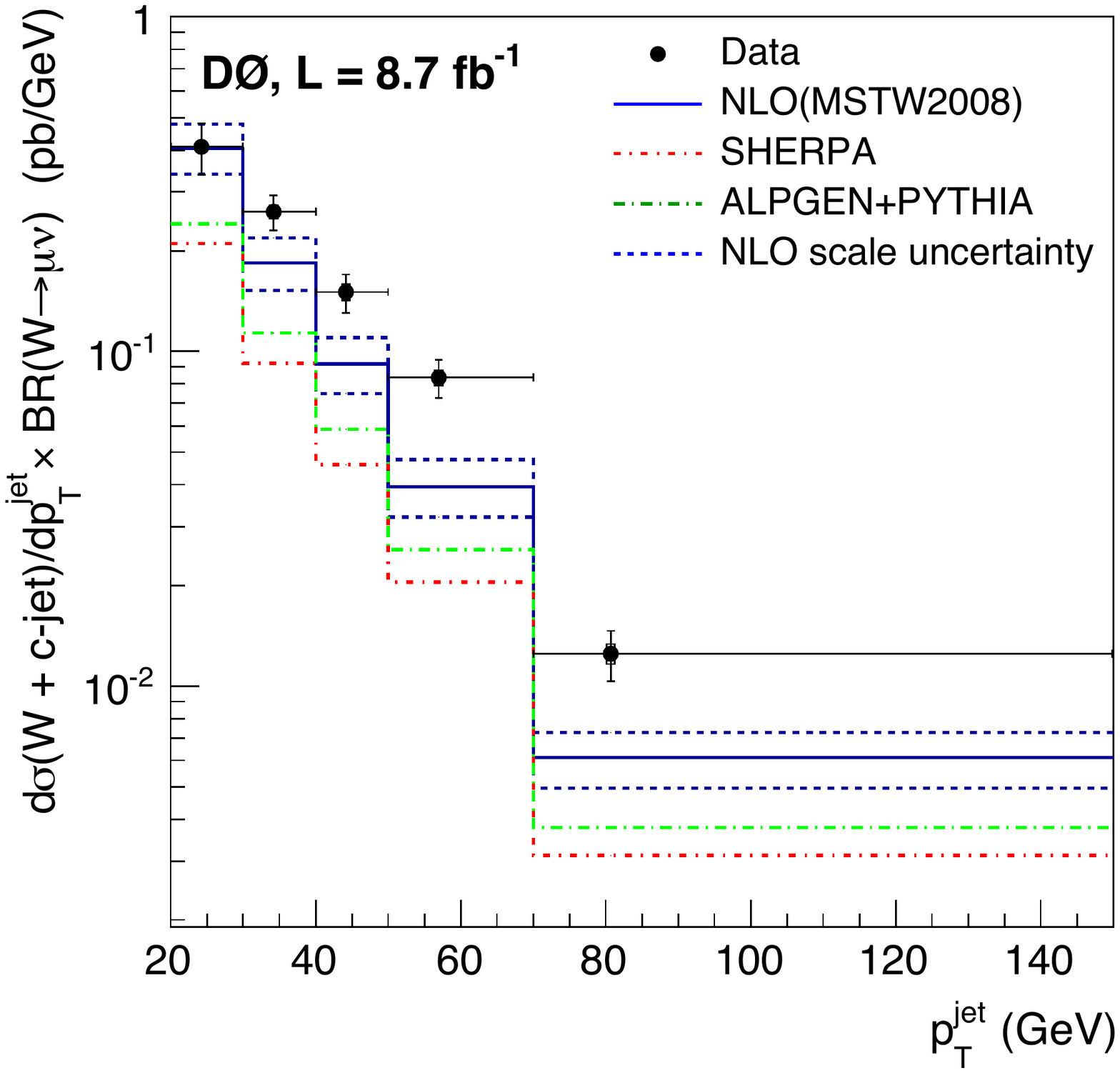}
\includegraphics[width=2.5in]{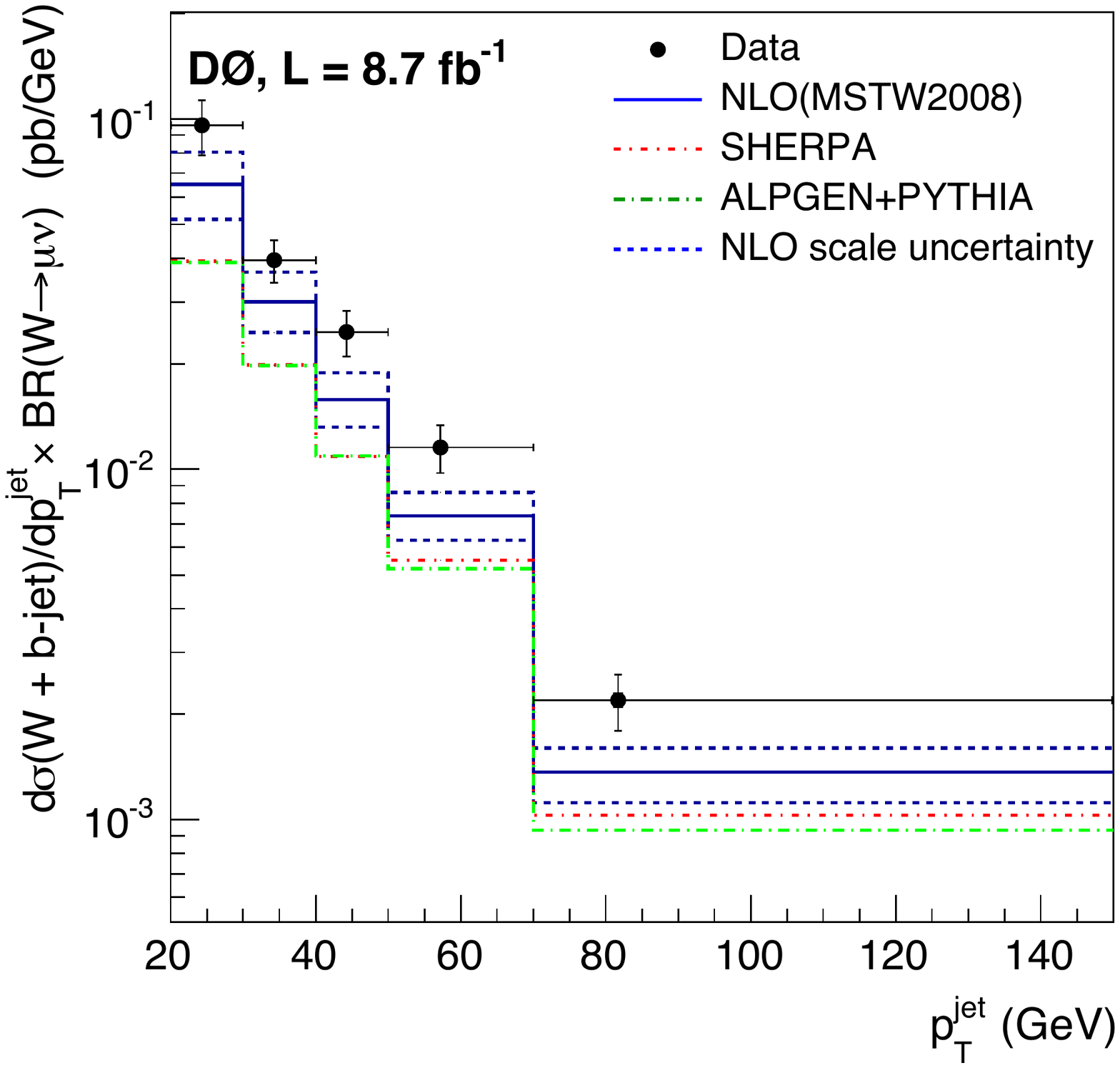}
\caption{Measured and predicted cross section for W$+$c quark production
(left) and W$+$b quark production (right) as a function of the transverse
momentum of the jet produced by the heavy quark.}
\label{fig:d0whf}
\end{figure}

CDF measured the cross section for prompt isolated central photon pair
production \cite{cdfdipho}. The measured cross section was studied as a
function of a large set of different kinematic variables, sensitive to
different details of the diphoton production mechanism, and compared with
a variety of state-of-the-art predictions, including parton-shower,
fixed-order, and analytically resummed calculations matched to fixed-order
NLO terms. Figure~\ref{fig:cdfdipho} shows a characteristic cross section
spectrum, differential in the azimuthal distance $\Delta\phi$ between the
two photons in the event, compared with various predictions. Only the
highest-order calculations, the full fixed-order NNLO calculation \cite{nnlo}
and the parton-shower {\sc sherpa} calculation with up to three additional
jets at the tree level, can reproduce reasonably well the entire spectrum.
Similar observations in diphoton production were later made by DZero
\cite{d0dipho}. This measurement was the first to test a full NNLO
calculation of a QCD process.

\begin{figure}[htb]
\centering
\includegraphics[width=5in]{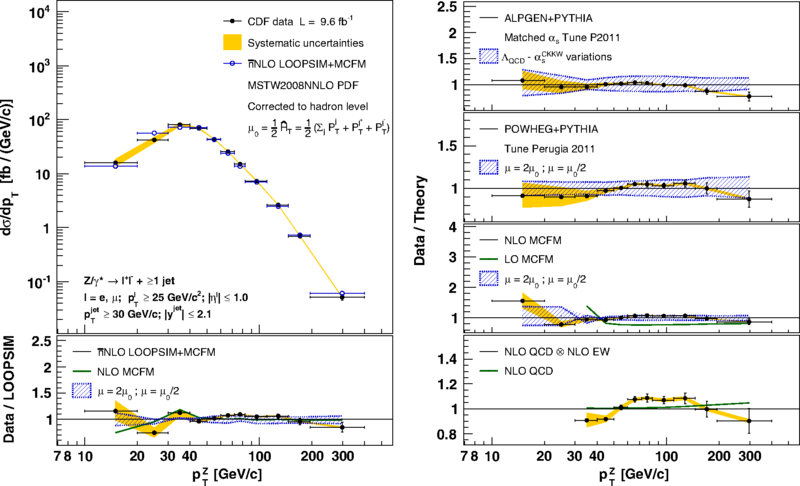}
\caption{Measured and predicted cross section for Z$+$jets production
as a function of the reconstructed Z boson transverse momentum (top left).
The bottom left and the right panels show the ratios of the measured
cross section to various predictions.}
\label{fig:cdfzjet}
\end{figure}

\section{Weak boson plus jets production}

CDF measured the cross section for W boson production in association with
light-flavor jets \cite{cdfwjet}, with the W boson reconstructed from the
leptonic decays W$\to$e$\nu_{\rm e}$ and W$\to$$\mu\nu_{\mu}$. In this
measurement, particular emphasis was given to the understanding of backgrounds
associated with W$+$jets production and the jet energy corrections required
to achieve a good agreement between data and simulations.
Figure~\ref{fig:cdfwjet} shows the measured and predicted cross section
as a function of the leading jet transverse energy (left window) and the
inclusive jet multiplcity (right window). The parton-shower calculation
with up to four jets at the tree level, multiplied by a K-factor of 1.4
to account for loop corrections, describes the data well within uncertainties.

\begin{figure}[htb]
\centering
\includegraphics[width=5in]{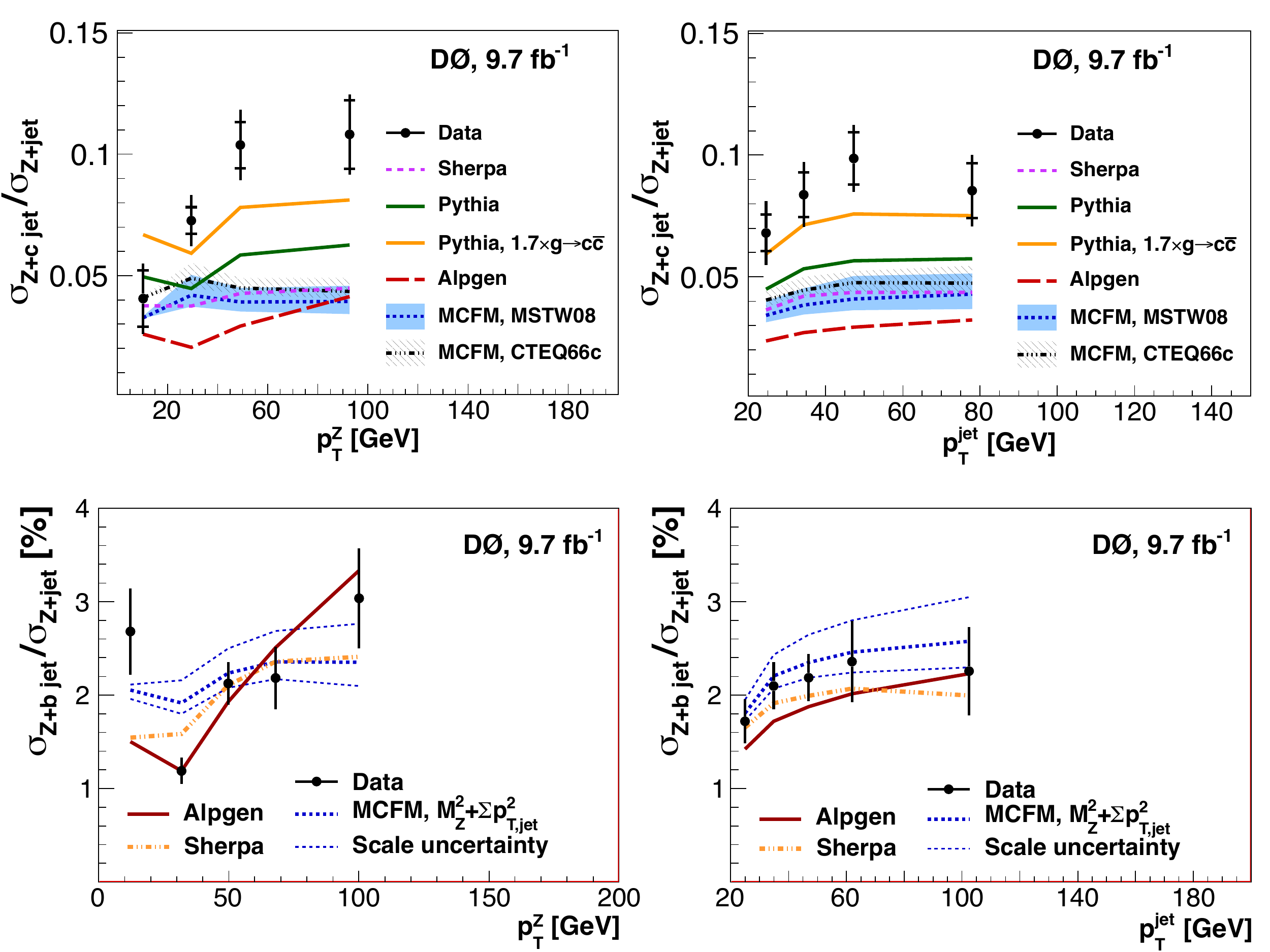}
\caption{Ratios of cross sections for Z$+$c (top) and Z$+$b quark production
(bottom) to the Z$+$jet production cross section as a function of the Z boson
transverse momentum (left) and of the jet transverse momentum (right).}
\label{fig:d0zhf}
\end{figure}

DZero measured the cross sections for the production of a W boson in
association with a heavy-flavor quark \cite{d0whf}, charm or bottom,
with the W boson recosntructed only from W$\to$$\mu\nu_{\mu}$ decays.
These processes are significant backgrounds in many searches for new physics
and precision SM measurements. The test of their description by theory is,
therefore, important. Figure~\ref{fig:d0whf} shows the measured and predicted
cross sections in the top panel and the corresonding ratios to the NLO
predictions in the bottom panel. Neither the tree-level parton-shower
calculations nor the fixed-order NLO calculation describe the data adequately.
All calculations underestimate the measured cross sections.

\begin{figure}[htb]
\centering
\includegraphics[width=3.5in]{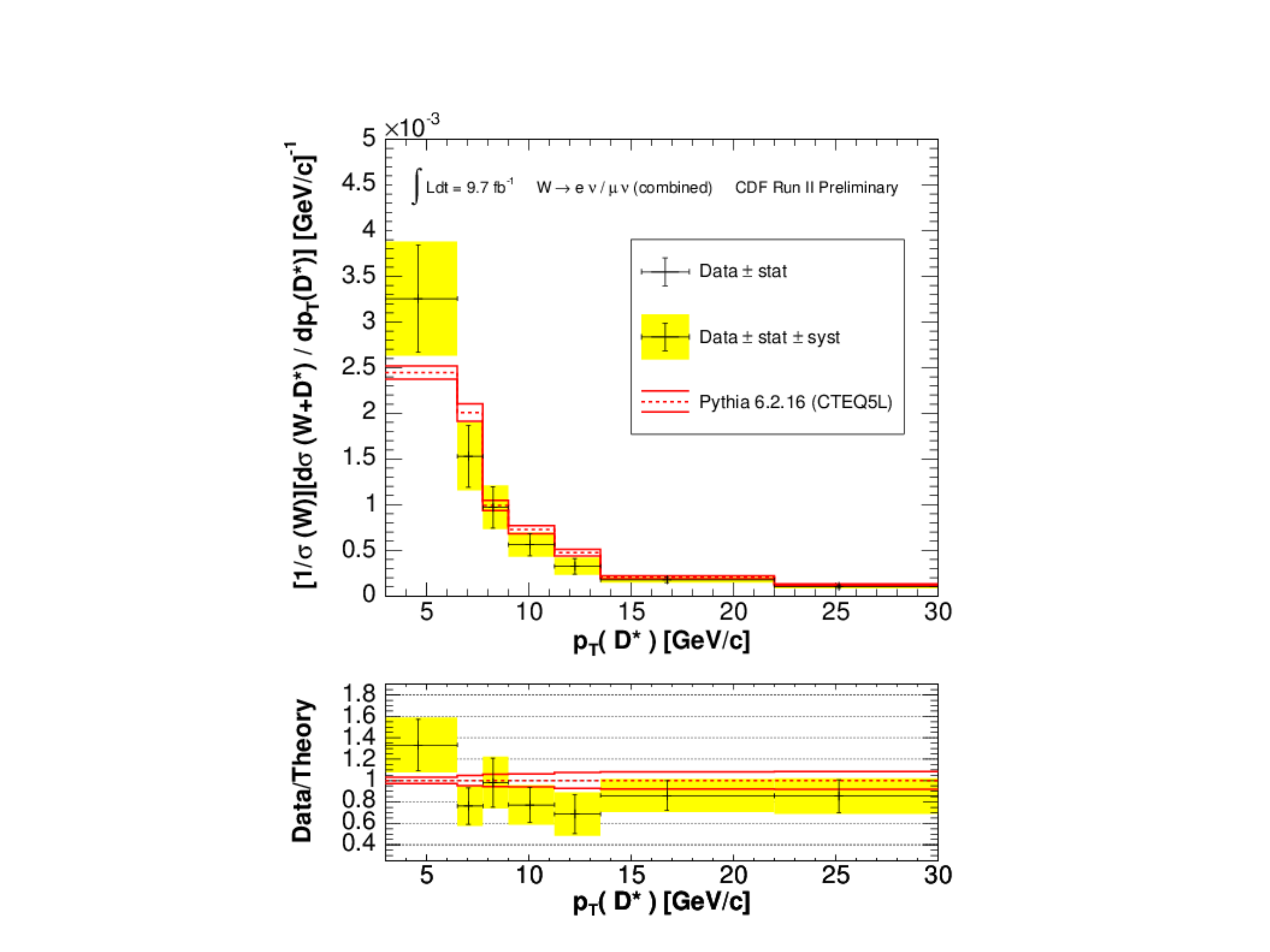}
\caption{Measured and predicted cross section for W$+$D$^*$ production
as a function of the D$^*$ meson transverse momentum. The lower panel
shows the ratio of the measurement to the prediction.}
\label{fig:cdfwd}
\end{figure}

CDF measured the cross section for the production of a Z boson in association
with light-flavor jets \cite{cdfzjet}. The Z boson was reconstructed from
the leptonic decays Z$\to$ee and Z$\to$$\mu\mu$. The Z$+$jets production
is also an important background in searches and precision measurements.
In this measurement, the cross section for various jet multiplicities,
differential in many kinematic variables sensitive to different aspects
of the production mechanism, was studied in detail and compared with a
variety of state-of-the-art calculations. These include tree-level and full
NLO matched to parton-shower calculations, parton-level calculations at
higher order, and fixed-order calculations at NLO both in the strong and
electroweak couplings. Figure~\ref{fig:cdfzjet} provides an example of the
cross section as a function of the reconstructed tranverse momentum of the
Z boson and the ratios of the measured cross section to various predictions.
In the overall, the calculations describe the data adequately, within
experimental and theoretical uncertainties.

\begin{table}[t]
\begin{center}
\begin{tabular}{lcc}
\hline\hline
 & $\Upsilon$$+$W & $\Upsilon$$+$Z \\ \hline
Expected limit (pb) & 5.6 & 13 \\
Observed limit (pb) & 5.6 & 21 \\
Run I observed limit (pb) & 93 & 101 \\ \hline\hline
\end{tabular}
\caption{Limits on $\Upsilon$$+$weak boson production total cross sections.}
\label{tab:cdfyz}
\end{center}
\end{table}

DZero measured the cross sections for Z boson production in association
with a heavy-flavor jet \cite{d0zhf}, originating from a charm or a bottom
quark, normalized to the cross section for Z boson production associated
with a light-flavor jet. The Z boson was reconstructed from the leptonic
decays Z$\to$ee and Z$\to$$\mu\mu$. Some uncertainties cancel out in the
cross section ratios. Figure~\ref{fig:d0zhf} shows the results compared
with various predictions as a function of the reconstructed transverse
momentum of the Z boson (left windows) and of the jet (right windows) for
a c-jet (top panel) and a b-jet (bottom panel). In general, the theory
describes reasonably well the Z$+$b-jet data, but not the Z$+$c-jet data.

\begin{figure}[htb]
\centering
\includegraphics[width=2.5in]{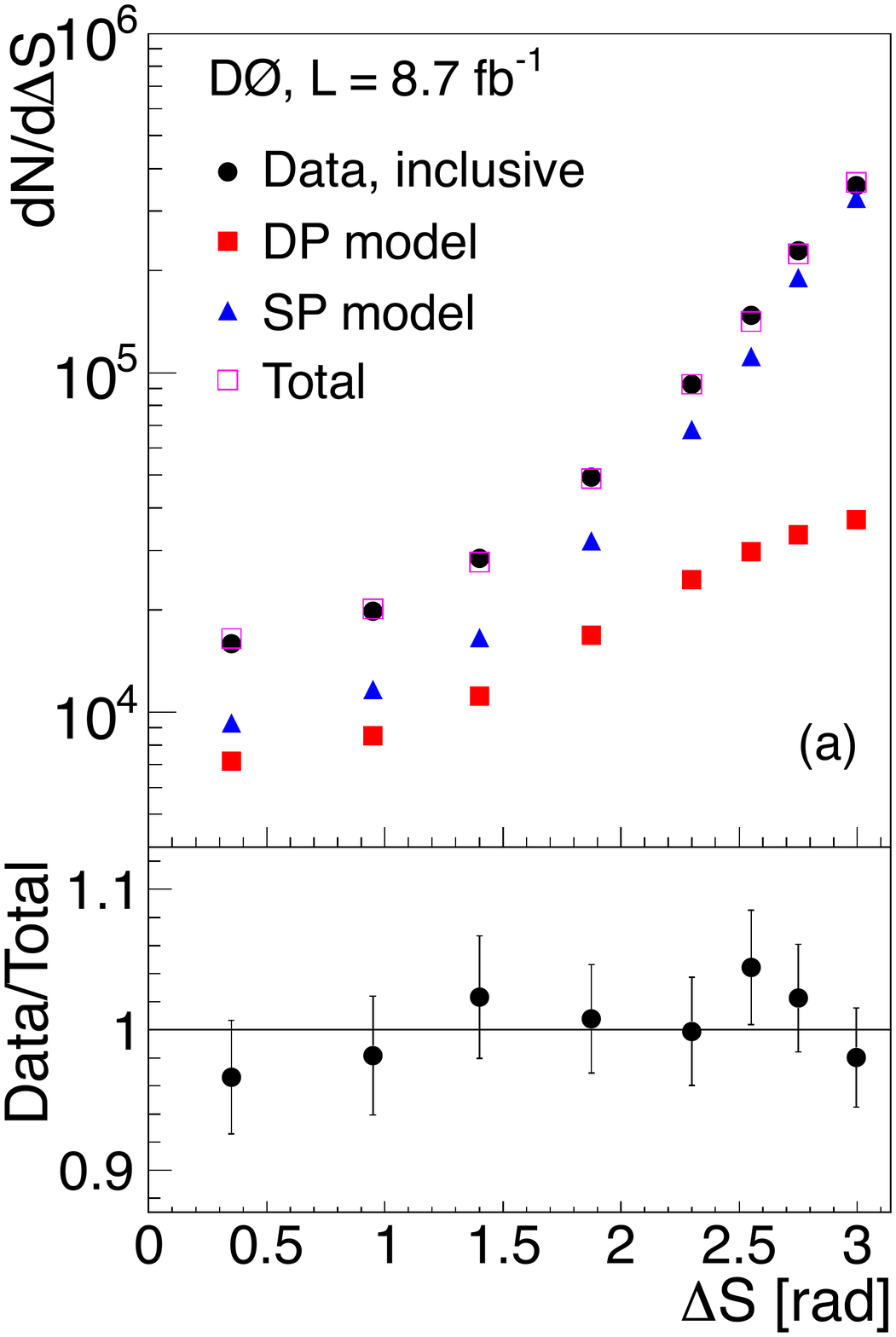}
\includegraphics[width=2.5in]{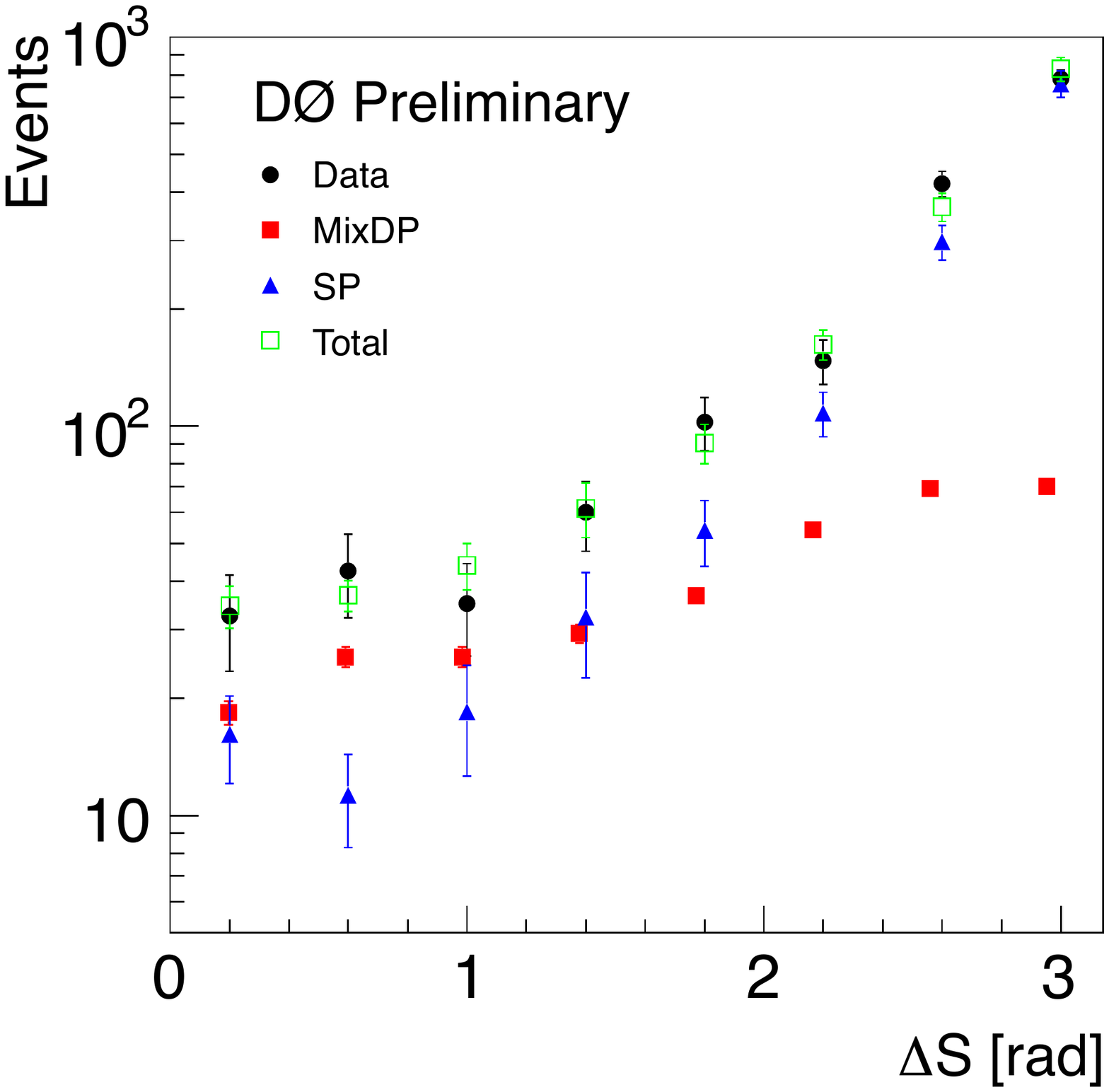}
\caption{Measured and predicted event distributions for photon$+$trijet
(left) and diphoton$+$dijet (right) production as a function of the
azimuthal separation between the leading and subleading pair of objects
in the 4-body final state. The bottom left panel shows the ratio of the
measured distribution to the total prediction.}
\label{fig:d0mpi}
\end{figure}

\section{Non-perturbative QCD processes}

A class of processes of interest to study is characterized by low momentum
transfers in interactions among high-momentum partons. Such processes are
important for the understanding of the role of non-perturbative QCD in
high-energy hadron collisions. They typically involve the production of
low-momentum isolated hadrons, where either the matrix element for the
formation of a QCD bound state (``long-range'' matrix element) in a hard
scattering process or the phenomenology of an entirely soft scattering
process (e.g. diffractive scattering, underlying event activity) plays
the dominant role.

\begin{figure}[htb]
\centering
\includegraphics[width=2.5in]{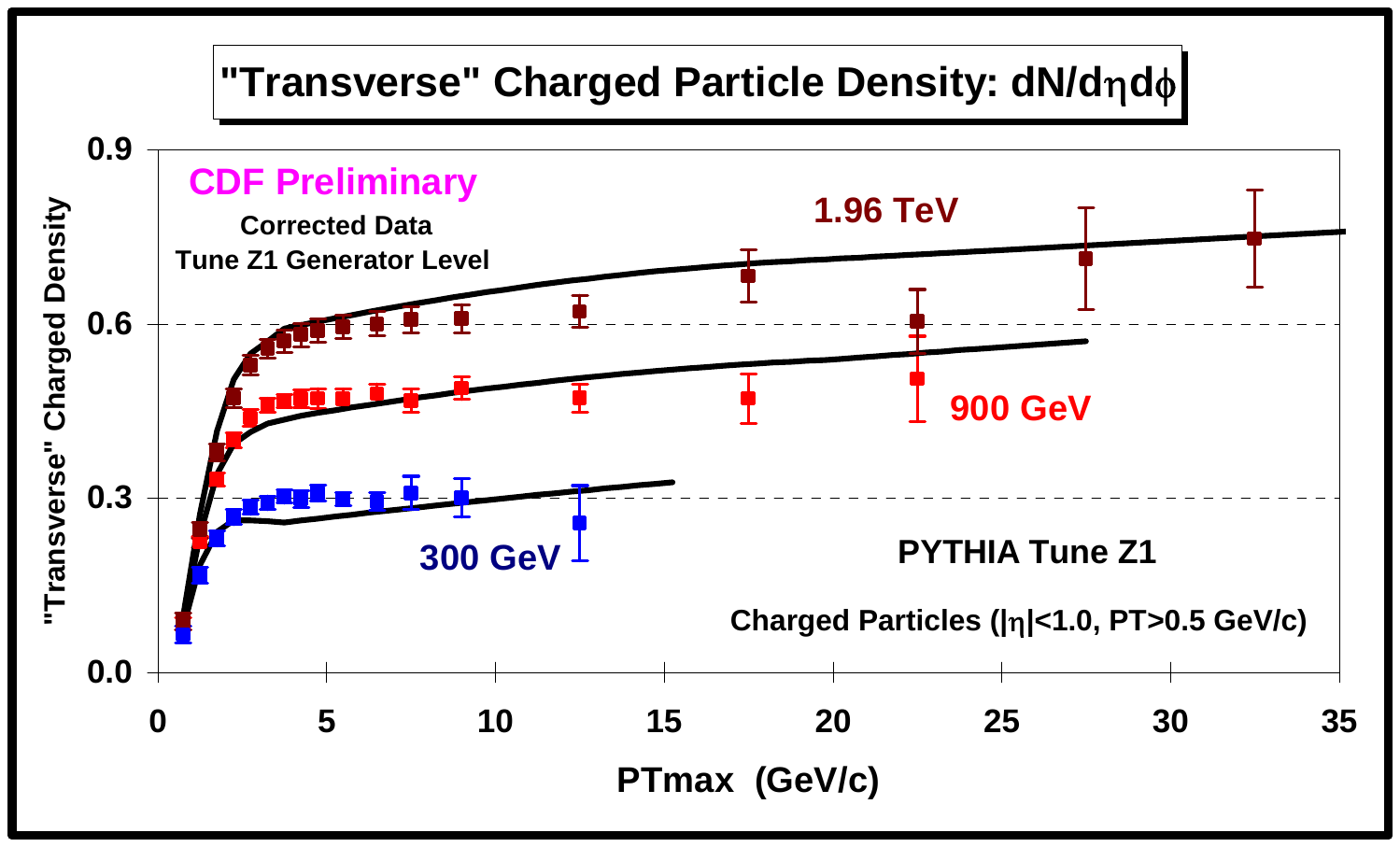}
\includegraphics[width=2.5in]{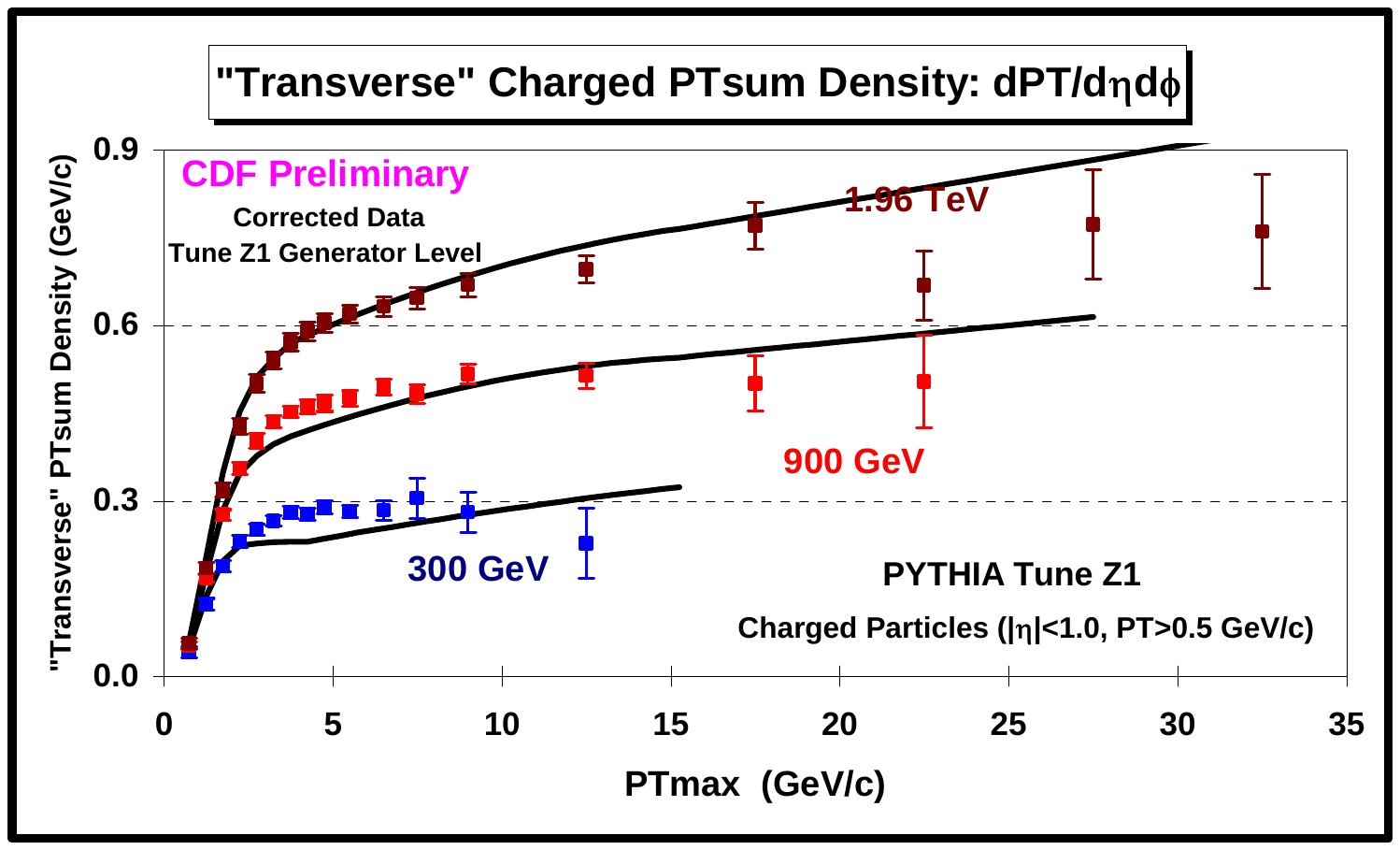}
\caption{Measured and predicted inclusive particle densities as a function
of the maximum single-object transverse momentum in the event.}
\label{fig:cdfue}
\end{figure}

CDF measured the cross section for the production of a leptonically decaying
weak boson, W or Z, in association with an $\Upsilon$ meson \cite{cdfwzy}
and with a D$^*$ meson \cite{cdfwd}. In each measurement, of interest is the
production mechanism of heavy flavor at low transverse momentum.
Table~\ref{tab:cdfyz} shows the observed and SM expected limits on the
$\Upsilon$$+$W and $\Upsilon$$+$Z production cross sections, compared with
the corresponding observed limits from Run I. Although the experiment did
not finally reach the statistical power to resolve the SM prediction from a
possible anomalous cross section due to new physics, the only so far existing
limits from CDF Run I are now improved by an order of magnitude.
Figure~\ref{fig:cdfwd} shows the measured and predicted cross section for
W$+$D$^*$ production as a function of the reconstructed transverse momentum
of the D$^*$ meson. The parton-shower calculation, with a LO matrix element
for W production and the D$^*$ production modeled through the showering and
hadronization prescriptions of the calculation, describes the data well
within uncertainties.

\begin{figure}[htb]
\centering
\includegraphics[width=2.5in]{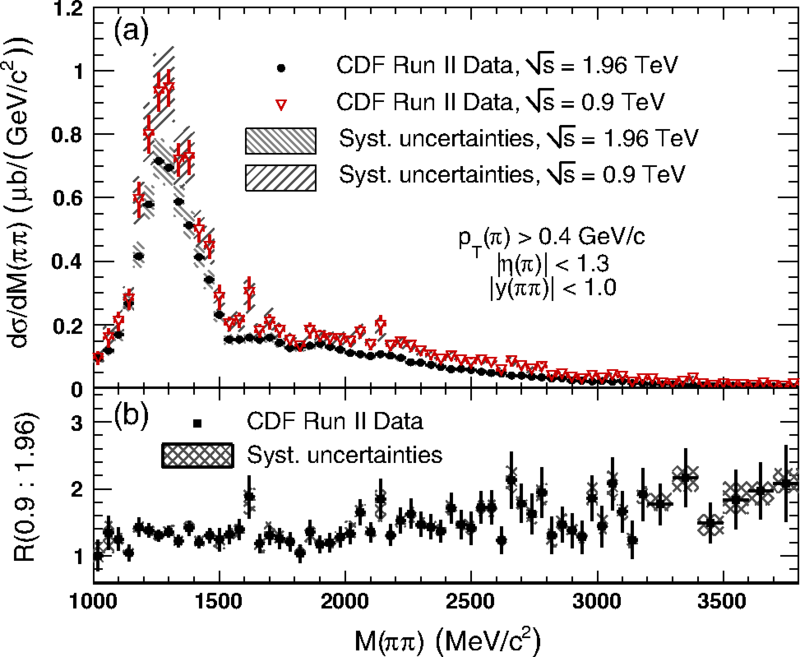}
\includegraphics[width=2.5in]{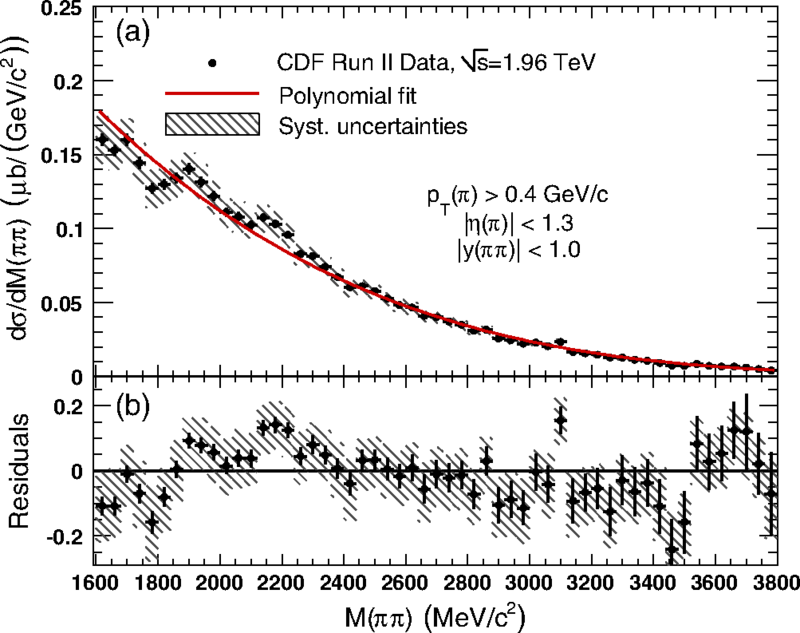}
\caption{Measured cross section for exclusive central pion pair production
at two collision energies as a function of the invariant mass of the pair
for the full mass range at both energies (left) and the high mass range
at the highest energy (right). The bottom left panel shows the difference
between the cross sections at the two energies. The bottom right panel shows
the residuals of the cross section at the highest energy from an empirical
polynomial fit.}
\label{fig:cdfdipi}
\end{figure}

DZero is conducting a suite of measurements aiming to study multiple parton
interactions in hadron collisions \cite{d0mpi}. The strategy is based on
analyses of events with 4-body final states, looking into decorrelations of
the two pairs of objects ordered in transverse momentum. The analyses look
at kinematic variables sensitive to these decorrelations. An example is the
azimuthal angle $\Delta$S in the plane normal to the colliding beams between
the transverse momentum vectors of the pair of the hardest (highest transverse
momentum) reconstructed objects and the pair of the softest (lowest transverse
momentum) reconstructed objects. For correlated pairs, this angle tends to be
large, as a result of the transverse momentum balance between two pairs
originating from a single parton (SP) interaction. For uncorrelated pairs,
it tends to be random, signifying that the two pairs originate from double
parton (DP) scattering, where the two parton interactions are nearly
independent. Figure~\ref{fig:d0mpi} shows the measured and predicted event
distributions for photon$+$3 jets events (left window) and for 2 photons$+$2
jets events (right window) as a function of $\Delta$S. The predictions agree
well with the data, showing that mutliple parton interactions are understood
in these processes.

CDF has performed a detailed program of underlying event (UE) studies.
The strategy of this program involves the measurement of inclusive
distributions of particles emerging in directions away from the transverse
momentum vectors of hard dijets, which are then used to tune phenomenological
UE models. In the last measurement \cite{cdfue}, the UE dependence on the
collision energy was studied using the full luminosity from all three Run II
collision energies of 300, 900, and 1960 GeV. Figure~\ref{fig:cdfue} shows
the measured and modeled densities of charged particles (left window) and
their summed transverse momentum (right window) for particles emitted away
from the jets. With the appropriate ``tune Z1'' of the {\sc pythia} event
generator \cite{pythia}, the UE model describes the measured density
distributions reasonably well within uncertainties.

Finally, CDF is pursuing a program of measurements of exclusive hadron
production in events with only an even number of low-momentum central
isolated tracks \cite{cdfdipi}. Such events are diffractive, characterized
by large gaps with no particles produced at large pseudorapidities, and they
are sensitive to t-channel double pomeron exchange. This program also exploits
the three Run II collision energies. Figure~\ref{fig:cdfdipi} shows the
invariant mass distribution of charged pion pairs in the full measured range
for 900 and 1960 GeV collision energy (left window) and in the high-mass
range for the 1960 collision energy (right window), where the statistics is
significant. The low-mass peak originates from the f$_2$(1270) and f$_0$(1370)
mesons decays, while more structure is visible in the high-mass tail. The
program probes a so far entirely unexplored region of soft QCD.

\section{Conclusions}

The Tevatron experiments, CDF and DZero, continue producing high-precision
results that provide stringent constraints of QCD calculations. The
experimental precision challenges the precision of NLO calculations in
most cases. NNLO calculations are generally needed to adequately describe
processes involving prompt photon production. W/Z$+$jets measurements,
in general, are successfully described by NLO calculations within experimental
and theoretical uncertainties, with the exception of W boson production
in association with heavy flavor quarks. Multiple parton interactions
are reasonably well understood. Long-range matrix elements and diffractive
physics are probed more precisely with the full luminosity data. The variety
of collision energies reached in the end of Run II allows for the study of
the energy dependence of non-perturbative QCD effects in certain cases.

\end{document}

%% file: econfmacros.tex
\def\beq{\begin{equation}}
\def\eeq#1{\label{#1}\end{equation}}
\def\eeqn{\end{equation}}

\def\beqa{\begin{eqnarray}}
\def\eeqa#1{\label{#1}\end{eqnarray}}
\def\eeqan{\end{eqnarray}}

\let\bar=\overbar

\def\Dslash{\not{\hbox{\kern-4pt $D$}}}
\def\dslash{\not{\hbox{\kern-2pt $\del$}}}

\def\msb{{\bar{\ssstyle M \kern -1pt S}}}

